\documentstyle[12pt,epsfig,cite]{article}

\newlength{\dinwidth}
\newlength{\dinmargin}
\def\titlepage{\clearpage%
\setcounter{footnote}{0}\setcounter{page}{1}%
\thispagestyle{empty}\pagestyle{plain}\pagenumbering{arabic}%
\kern1mm\begin{center}
\end{center}
\vskip3mm\normalsize}
\def\docnum#1{\hbox to \hsize{\hskip123mm\hbox{#1}\hss}}
\def\date#1{\edef\@temp{#1}\ifx\@temp\@empty\def\@temp{\today}\fi
\hbox to \hsize{\hskip123mm\hbox{\@temp}\hss}}
\def\title#1{\vskip 0.8in plus 2in\begin{center}%
{\Large\bf#1\par}\vskip1.5em\end{center}\vskip 1in}
\def\@makefnmark{\hbox{$^{\@thefnmark)}$}}
\def\author#1{
\setcounter{footnote}{0}\def\@currentlabel{}%
\begingroup\def\thefootnote{\arabic{footnote}}
\def\@makefnmark{\hbox{$^{\@thefnmark)}$}}
\global\@topnum\z@ \large\begin{center}{\lineskip.5em
\begin{tabular}[t]{c}#1\end{tabular}\par}
\end{center}\par\vskip1.5em\@thanks\endgroup}

\def\abstract{\vskip0.8in plus 3in\begin{center}{\large\bf
Abstract}\end{center}\quotation}

\newcommand{\QG}  {{\bf Q}}
\newcommand{\PG}  {{\bf \phi}}
\newcommand{\qi}  {{\bf q}_i}
\newcommand{\qj}  {{\bf q}_j}

\begin{document}
\begin{titlepage}

\flushright{DFF 234/10/1995}
\flushright{October 1995}
\title{A thermodynamical model of hadron
production in $e^+e^-$ collisions}
\vspace{1cm}
\centerline{\large{F. Becattini}}
\vspace{0.5cm}
\centerline{\it{Universit\`a di Firenze and INFN Sezione di Firenze}}
\centerline{\it{Largo E. Fermi 2, I-50125 Firenze}}
\centerline{e-mail: becattini@vaxfi.fi.infn.it}
\vspace{1.5cm}

\begin{abstract}
The hadron production in $e^+e^-$ collisions is studied assuming
that particles originate in a hadron gas at thermal and chemical
equilibrium. The parameters of the hadron gas are determined with
a fit to the average multiplicities of various hadron species
measured at LEP and PEP-PETRA colliders. An impressive agreement
is found between the predictions of this model and data for almost all
particles over a range of production rate of four orders of magnitude.
The
temperature values found at the centre of mass energies of LEP and
PEP-PETRA are around $165$ MeV.
\end{abstract}
\vspace{2cm}
\centerline{\it{To be published in the Proceedings of the }}
\centerline{\it{XXV International Symposium on Multiparticle Dynamics}}

\end{titlepage}

\section{Introduction}

Hadron production in $e^+e^-$ collisions at high energy is
generally believed
to be the result of a two-stage process: a parton shower
generated by the
$q \bar q$ pair emerging from the annihilation, and a fragmentation of
the partons into observable hadrons. The former process is hard and
can be described
by perturbative QCD, whereas the latter is soft and not calculable
with a  perturbative approach.\\
Several phenomenological models aimed at describing quantitatively the
fragmentation process have been developed; amongst them, the most
popular are those
implemented in the Monte Carlo generators JETSET \cite{jet} and
HERWIG \cite{her},
using the concepts of string and cluster fragmentation respectively.
The main unsatisfactory feature of these models, as far as the
relative production
rates of the hadron species is concerned, is the large number
of free parameters
required in order to correctly reproduce experimental data. As
a consequence,
those models have a rather poor predictive power. Empirical
regularity in hadron
production rates have been
observed in ref. \cite{str,let} although some non-physical
assumptions were
made about meson quantum numbers and masses. The most
interesting result of
those observations is that the very same exponential
behaviour of the hadron
production function is present both at LEP ($\sqrt{s}=91.2$ GeV)
and PEP-PETRA
($\sqrt{s} = 29 \div 35$ GeV) centre of mass energies.\\
We introduce a thermodynamical approach to the problem of
hadronization \cite{beca},
postulating the existence of a hadron gas in thermodynamical
equilibrium before
the hadrons themselves decouple (freezing-out) and decay giving
rise to observable
particles in the detector. We show that this model is able to
fit almost all inclusive
rates measured so far at LEP and PEP-PETRA colliders in a natural
way by using only
three parameters.
Two of them are the basic parameters of the hadron gas description,
namely its
temperature $T$ and its volume $V$; the third one, $\gamma_s$ is a
parameter
describing the partial strangeness chemical equilibrium,
which has been used already in some analysis of hadron
production in heavy-ion
collisions \cite{gam}. The only required inputs to determine
the particle yields
are the mass, the spin and the quantum numbers such as baryon
number, strangeness, charm and beauty.

\section{Hadron gas scheme}

Most hadronic events in high energy $e^+e^-$ annihilations
are two-jet events.
Here we assume that each jet represents a hadron gas phase
in complete thermodynamical
equilibrium just before the freezing-out time and that the
number of such phases in
hadronic events is always two, that means neglecting multi-jets
events.
One can then describe a jet as an object defined by thermodynamical
and mechanical
quantities such as the temperature and the volume in its rest frame
and the
Lorentz-boost factor $\gamma$. As far as chemical equilibrium is
concerned, we
assume that, in general, a jet has quantum numbers related to
those of the
parent quark. As the two jets must be colourless and do not
have fractional baryon
numbers, sharings of one or several quark-antiquark pairs
should occur.
It seems reasonable to assume that most pairs are of either
$u \bar u$ or
$d \bar d$ type which are the lightest ones. We allow also
$s \bar s$ sharing
in the inter-jet interaction and non-vanishing baryon number
for each jet
provided that the baryon number and strangeness of the whole
system are zero.
Morevoer, we assume that each jet keeps the charm and beauty
of the parent quark,
i.e. no heavy quark pairs are exchanged.
Adequate tools to deal with such a problem in a statistical
mechanics framework
are the canonical partition functions of systems with internal
symmetries. The
partition function of a system which transforms under the
irreducible representation
$D^\nu$ of a symmetry group $G$ can be expressed as \cite{zf1,zf2}:

 \begin{equation}
     Z = \frac{d_\nu}{M(G)} \, \int\, d\mu(g)\,
     \chi_{\nu}(g^{-1}) \, {\cal Z}(g) \; ,
 \end{equation}
where $\mu(g)$ is the group measure, $M(G)=\int d\mu(g)$, $d_\nu$
is the dimension
of $D^\nu$ and $\chi_{\nu}(g)$ is the character of $D^\nu$, namely
$\chi_{\nu}(g)={\rm tr} \,(D^\nu(g))$. The function ${\cal Z}$
is defined as:

 \begin{equation}
     {\cal Z}(g) = {\rm tr} \, ( e^{- \beta E}
     e^{-i \sum \phi_i Q_i} ) \; ,
 \end{equation}
where the $\phi_i$'s are the parameters of the group and
$Q_i$ its generators.
The exponential factor is the usual canonical distribution
with $\beta$ as the
inverse temperature and $E$ as the energy of the system,
both calculated in
its rest frame.\\
In the present case the symmetry group is $U(1)^4$, each
$U(1)$ corresponding
to the conservation of baryon number $N$, strangeness $S$,
charm $C$ and
beauty $B$ respectively. The Eq. (1) becomes:

\begin{equation}
     Z(\QG)=\frac{1}{(2\pi)^4}
  \int_{0}^{2\pi}\!\!\!\!
  \int_{0}^{2\pi}\!\!\!\!
  \int_{0}^{2\pi}\!\!\!\!
  \int_{0}^{2\pi} \, d^4 \phi \,\, e^{i \QG \cdot \PG}\,
  {\cal Z}(\PG) \; .
\end{equation}
$\QG = (N,S,C,B)$ is a four dimensional vector with integer
components representing the quantum numbers of a jet and
$\PG =(\phi_1,\phi_2,\phi_3,\phi_4)$, each $\phi_i$ being the
parameter of a $U(1)$ group.\\
According to the assumptions made for the quantum numbers of
the system, the partition function of the whole system can
be written as:
\begin{equation}
            \hat{Z} = \sum_{N=0}^1 \sum_{S=-2}^2 Z(\QG) Z(-\QG) \; ,
\end{equation}
where $Z(\QG)$ are the partition functions of the single jets
and the sum
runs over baryon number $N$ and strangeness $S$. Further
constraints are
implicitly introduced in Eq. (4), namely a maximum allowed
baryon number 1
and a maximum allowed strangeness 2. Moreover, in case of $e^+ e^-
\rightarrow s \bar s$
events the term with $S=0$ is removed from the sum. These assumptions
define what has been called correlated jet scheme \cite{beca}.
For a gas of $N_B$ boson species and $N_F$ fermion species,
the partition function of a single jet becomes:

\begin{eqnarray}
    Z(\QG ) & = & \frac{1}{(2\pi)^4} \int\, d^4 \phi
     \,\, e^{i \QG \cdot \PG}
     \exp \, \{ {\sum_{j=1}^{N_B}\sum_k \,\log \,
     (1-e^{-\beta \varepsilon_{k} -i
     \qj \cdot \PG})^{-1}} + \nonumber \\
   && + {\sum_{j=1}^{N_F}\sum_k \,\log \,
   (1+e^{-\beta \varepsilon_{k} -i
     \qj \cdot \PG})}\} \; ,
\end{eqnarray}
where $k$ labels all available states of phase space for the
$j^{th}$ particle,
$\varepsilon_k$ is its energy and $\qj = (N_j,S_j,C_j,B_j)$.
Since the total number of each hadron in the gas
is not determined, all chemical potentials must vanish.
 From Eq. (4) we are able to estimate the average
number per jet of any hadronic
particle assigning each of them a fictitious
fugacity $\lambda_i$ which multiplies
the $e^{-\beta  \varepsilon_{k}}$ factors and deriving:

\begin{equation}
     <n_i> = \lambda_i \frac {\partial
      \log {\hat Z}}{\partial \lambda_i} |_{\lambda_i=1} \; .
\end{equation}
The sum over the phase space, for a continuous level density becomes:

\begin{equation}
 \sum_k \longrightarrow (2J+1) \frac {V}{(2\pi)^3} \int d^3 p \; ,
\end{equation}
where $J$ is the spin of the particle and $V$ the volume of
the hadron gas phase.\\
One is left with complicated integrals in Eq. (5),(6) but,
if $T \sim {\cal O}(100)$
MeV, as we can argue from the QCD soft scale, the exponential
factors are expected to be very small for all particles but
pions, so that

\begin{eqnarray}
  && \log \, (1\pm e^{-\sqrt{p^2+m^2_i}/T -i
     \qi \cdot \PG })^{\pm 1} \simeq e^{-\sqrt{p^2+m^2_i}/T -i
     \qi \cdot \PG } \nonumber \\
  && \frac {1}{e^{\sqrt{p^2+m^2_i}/T + i \qi \cdot \PG} \pm 1}
 \simeq e^{-\sqrt{p^2+m^2_i}/T - i \qi \cdot \PG}
\end{eqnarray}
are very good approximations. This corresponds to the Boltzmann
limit of Fermi and
Bose statistics. It can be shown then that the average number
of particles $i$ produced per jet is
\begin{equation}
  <n_i> = z_i \frac {Z (\QG - \qi)}{Z(\QG)} \; ,
\end{equation}
whereas for pions

\begin{equation}
  <n_i> = \frac {V}{(2\pi)^3} \int d^3 p \,\,
  \frac {1}{e^{\sqrt{p^2+m^2_i}/T}
  - 1}  \; ,
\end{equation}
being

\begin{equation}
  z_i = (2J_i+1) \frac {V}{(2\pi)^3} \int d^3 p \,\,
  e^{-\sqrt{p^2+m^2_i}/T} =
 (2J_i+1)\frac{VT}{2\pi^2} m^2_i K_2 (\frac{m_i}{T}) \; .
\end{equation}
$K_2$ is the modified Bessel's function of order 2.\\
The factor $Z(\QG - \qi)/Z(\QG)$ suppresses or enhances the
thermal production rate of
the particle according to its quantum numbers and the quantum
numbers of the jet
(see also Fig. 1).

\section{Fit of LEP and PEP-PETRA data}

The volume and the temperature of the two phases
at the decoupling time
are determined by a fit to the available data on hadron
inclusive production.
The volume and the temperature are assumed to be equal
for both jets and also equal
for different kinds of primary quarks. A parameter
$\gamma_s < 1$ is introduced
in order to take into
account a non-complete strange chemical equilibrium;
if a hadron contains $n$
strange valence quarks, its production rate is reduced
by a factor $\gamma_s^n$. This
reduction applies also to neutral mesons such as $\eta$,
$\eta'$, $\phi$, $\omega$
according to the fraction of $s\bar s$ content in the meson itself.\\
The hadron rates calculation proceeds through two steps;
in the first one
a primary number of hadrons emitted from the thermal source
is calculated
according to the formulae of previous section; in the second,
all decays of these
primary hadrons are performed according to known decay modes
and branching ratios.
The overall production of each particle is then computed by
adding up the fraction
stemming from decays of heavier particles to the primary one.
The decay chain stops
when $\pi$, $K$, $K_L^0$, $\mu$ or stable particles are reached,
in order to match
the numbers of production rates provided by experiments
at $e^+e^-$ colliders,
including all decay products of particles with $c\tau < 10$ cm.
All hadrons included in JETSET 7.4 tables have been considered,
with the quoted
masses and widths. All other light flavoured resonances up to a
mass of 1.7 GeV
have been included with masses, widths and branching ratios
quoted by PDG \cite{ppr}.
The mass of resonances with $\Gamma > 1$ MeV has been distributed
according to a
relativistic Breit-Wigner function within $\pm 2\Gamma$ from the
central
value. Experimental production rates used in the fit at
$\sqrt{s} = 91.2$ GeV \cite{dean}
and at $\sqrt{s} = 29 \div 35$ GeV \cite{ppr} are weighted
averages of all available
measurements.\\
The results of the fit are summarized in Tables 1, 2 and Figs. 2, 3.
The errors on the parameters $T$, $V$ and $\gamma_s$ estimated
from the fit
include also uncertainties on input data such as branching
ratios, masses and
widths of particles involved in the fit. The fit procedure
is described more in
detail in ref. \cite{beca}.\\
An excellent agreement between
experimental and calculated production rate values
is achieved. A significative
deviation is observed at $\sqrt{s} =91.2$ GeV for
the $\Sigma^*$ production rate whose experimental value
is still controversial
because of the large disagreement between different
experiments \cite{sigma}.\\
By using as input data the fitted $T$, $V$ and
$\gamma_s$ values, the model also
yields a definite prediction for the relative
production rates of heavy flavoured
hadrons once $R_c = \sigma(e^+e^- \rightarrow c
\bar c)/\sigma(e^+e^- \rightarrow
hadrons)$ and $R_b = \sigma(e^+e^- \rightarrow b
\bar b)/\sigma(e^+e^- \rightarrow
hadrons)$ are known.\\
In Table 3 the predicted multiplicities of some of the
lightest heavy flavoured hadrons are quoted together with
corresponding
experimental vaues \cite{hf}. A very good agreement with
data is found.
Two main sources of systematic uncertainties have been
considered. The
first one is simply a possible systematic error in the
parameters fit which has
been estimated by excluding the data points deviating
the most from theoretical values and repeating the fit
itself. The second one is related to possible
different values of $V$ as a function of the primary quark mass. This
has been taken into account by introducing two dimensionless variables
$x_c, x_b \in [0,1]$
such that $x_c V$ is the volume in $e^+e^- \rightarrow c \bar c$
and $x_b V$ in
$Z \rightarrow b \bar b$ events, whereas
$V(1-x_c R_c - x_b R_b)/(R_u+R_d+R_s)$ is
the volume for $Z \rightarrow q \bar q$
where $q$ is a light quark.
The measurement of the averaged charged
multiplicity in $e^+e^- \rightarrow c \bar c$
and $e^+e^- \rightarrow b \bar b$ \cite{bclep,bclow}
determine $x_c = 0.88\pm0.03$
and $x_b=0.70\pm0.016$ at
$\sqrt{s} = 91.2 $ GeV and $x_c = 0.89\pm0.12$ and
$x_b=0.47\pm0.15$ at
$\sqrt{s} = 29 \div 35 $ GeV. Inserting those values
in the fit, it is found that new
fitted parameters are almost unchanged.
Other systematic effects related to
uncertainties on $R_q$ are found to be
negligible. The variations of the
fitted parameters due to those mentioned effects
are summarized in Table 4.\\
Another important issue is
the independence of the fitted parameters, in particular
$T$ and $\gamma_s$, from the
neglected part of the hadronic mass spectrum,
i.e. light-flavoured resonances
with mass $> 1.7$ GeV. This has been tested by moving
the cut-off point down to 1.3 GeV
by steps of 0.1 GeV. The new values are found to be
consistent with the main quoted values
within the fit error, as shown in Fig. 4.
Morevoer, the number of primary hadrons
stabilizes around 17 as the cut-off points
increases. This indicates that missing
heavy resonances production is negligible
and should not affect significantly the fit.\\
One of the most relevant results is that
temperatures values at both centre of mass
energies turn out to be very close. If we
add in quadrature the systematic uncertainties
quoted in Table 4 to the fits error quoted in Table 1 we get:
\begin{eqnarray}
 T_{LEP} & = & 162.8\pm2.9 \\
 T_{PEP} & = & 169.5\pm3.7 \; .
\end{eqnarray}
The $\chi^2$ of the weighted average of
those values is 1.8, which indicates a good
degree of compatibility. Also, it should be
taken into account that the temperature value
of PEP-PETRA might be affected by an additional
systematic error due to having averaged production rates
measurements spread between 29 and 35 GeV.

\section{Conclusions}

The problem of the relative production of
hadrons in $e^+ e^-$ collisions is
treated with a thermodynamical approach
postulating that hadronic jets must be
identified with hadron gas phases in
thermodynamical equilibrium before the hadrons
themselves decouple and subsequently
decay into lighter particles. Nevertheless,
since the particle production depends
almost linearly on the volume, the occurrence
of additional jets, i.e. phases, in the
same event should affect the relative hadron
production rates very mildly provided
that the temperature of the additional jets
is the same.\\
It has been shown that this model is able to fit very well
the average multiplicities of light
hadrons per hadronic event observed at LEP
($\sqrt{s}$=91.2 GeV) and PEP-PETRA
($\sqrt{s}=29 \div 35$ GeV) for a variation
of production rate by four orders of
magnitude, from $\pi$ to $\Omega$.\\
The predicted rates of heavy flavoured
hadrons are also in good agreement with data.
Only three parameters are required in
order to reproduce data correctly,
namely the temperature and the volume of the
hadron gas and a parameter $\gamma_s \in [0,1]$
allowing an incomplete strange chemical
equilibrium. The temperature values at
$\sqrt{s} = 91.2$ GeV and $\sqrt{s} =29 \div 35$ GeV,
determined with the fit, are very similar and indicate a constant
hadronization temperature. The thermalization
of the system could be a characteristic
of the quark-hadron transition, brought about by strong interactions.

\section*{Acknowledgements}

I wish to express my gratitude to A.
Giovannini and W. Kittel for their
keen interest in this work. I would
like to thank all the organizers for
having provided a favorable climate for the conference.

\newpage


\newpage

\section*{Figure captions}

\begin{itemize}

\item[\rm Figure 1]
Behaviour of the jet partition function $Z$ as a function of
baryon number $N$ for
$(S, C, B) = 0$ (left) and as a
function of the strangeness $S$ for $(N, C, B) = 0$ (right),
for $T=163$ MeV and $V=21$ Fm$^3$.
The suppression factors $Z(\QG-\qi)/Z(\QG)$
are smaller for increasing $N$ and $S$.

\item[\rm Figure 2]
Fit of hadron production measured at LEP.
Above: the solid line connects fitted
values while data are shown as black dots.
Below: fluctuations of measured points
from the fitted values (solid line) in
standard deviation units. Error bars
include also contributions from uncertainty
on masses, widths and branching
ratios of hadrons.

\item[\rm Figure 3]
Fit of hadron production measured at PEP-PETRA.
Above: the solid line connects fitted
values while data are shown as black dots. Below:
fluctuations of measured points
from the fitted values (solid line) in standard
deviation units. Error bars
include also contributions from uncertainty on
masses, widths and branching
ratios of hadrons.

\item[\rm Figure 4]
Fitted temperature $T$, volume $V$, $\gamma_s$ and number of primary
hadrons at $\sqrt{s} = 91.2$ GeV as a function of the light flavored
resonance  cut-off mass.

\end{itemize}

\newpage

\section*{Tables}

\begin{table}[ht]
   \begin{center}
   \begin{tabular}{| c || c | c |}
      \hline
  {\bf Parameters}&$\sqrt{s} =91.2$ GeV
  &$\sqrt{s} = 29 \div 35$ GeV  \\ \hline\hline
Temperature(MeV)  &  162.9$\pm$2.1
&  169.3$\pm$3.5             \\ \hline
 Volume(Fm$^3$)    &  21.4$\pm$1.9
 &  9.3$\pm$1.4               \\ \hline
$\gamma_s$        &  0.696$\pm$0.027
&  0.811$\pm$0.046           \\ \hline
$\chi^2/dof$      &  38.6/17
&  29.3/12                   \\ \hline
   \end{tabular}
\caption[]{Fit results.}
\end{center}
\end{table}

\begin{table}
   \begin{center}
   \begin{tabular}{| c || c | c || c | c |}       \hline
  & \multicolumn{2}{|c|}{$\sqrt{s} = 91.2$ GeV}
      & \multicolumn{2}{|c|}{$\sqrt{s} = 29 \div 35$ GeV} \\ \hline
{\bf Hadrons}   &  Calculated  & Measured
&    Calculated  &   Measured             \\ \hline
$\pi^0$         & 10.03        & $9.19\pm0.73\pm0.37$
    &   6.186       & $5.6\pm0.3\pm0.22$      \\ \hline
$\pi^+$         &  8.952       & $8.53\pm0.22\pm0.34$
    &   5.415       & $5.15\pm0.2\pm0.21$     \\ \hline
$K^+$           &  1.065       &$1.18\pm0.064\pm0.013$
    &   0.750       & $0.74\pm0.045\pm0.008$  \\ \hline
$K^0$           &  1.010       &$1.006\pm0.016\pm0.013$
    &   0.704       & $0.74\pm0.035\pm0.008$  \\ \hline
$\eta$          &  0.866       & $0.94\pm0.11\pm0.11$
    &   0.535       & $0.61\pm0.07\pm0.0056$  \\ \hline
$\omega$        &  0.939       & $1.11\pm0.14\pm0.011$
    &   0.544       &   --                    \\ \hline
$\rho^0$        &  1.186       & $1.29\pm0.13\pm0.17$
    &   0.704       & $0.81\pm0.08\pm0.11$    \\ \hline
$K^{*+}$        &  0.350       &$0.357\pm0.034\pm0.089$
    &   0.246       & $0.32\pm0.025\pm0.006$  \\ \hline
$K^{*0}$        &  0.344       &$0.379\pm0.020\pm0.089$
    &   0.241       & $0.28\pm0.03\pm0.006$   \\ \hline
$p$             &  0.534       &$0.489\pm0.05\pm0.041$
    &   0.293       & $0.32\pm0.025\pm0.015$  \\ \hline
${\eta}^{\prime}$        &  0.0993      &$0.22\pm0.07\pm0.0002$
     &   0.0702      & $0.26\pm0.1\pm0.00007$  \\ \hline
$\phi$          &  0.123       &$0.107\pm0.009\pm0.002$
  &   0.106       & $0.085\pm0.011\pm0.002$ \\ \hline
$\Lambda$       &  0.162       &$0.185\pm0.007\pm0.007$
  &   0.104       & $0.103\pm0.005\pm0.005$ \\ \hline
$(\Sigma^++\Sigma^-)/2$& 0.0366&$0.044\pm0.007\pm0.0035$
  &   0.0226      &   --                    \\ \hline
$\Sigma^0$      & 0.0383       &$0.036\pm0.007\pm0.0035$
  &   0.0243      &   --                    \\ \hline
$\Delta^{++}$   & 0.0925       &$0.062\pm0.032\pm0.011$
  &   0.0488      &   --                    \\ \hline
$\Xi^-$         & 0.0122       &$0.0128\pm0.0007\pm0.0004$
  &   0.00893     &$0.0088\pm0.0014\pm0.0003$\\ \hline
$(\Sigma^{+*}+\Sigma^{*-})/2$&0.0192&$0.011\pm0.002\pm0.0014$&
 0.0111      &$0.0085\pm0.002\pm0.0015$ \\ \hline
$\Xi^{*0}$      & 0.00421      &$0.0031\pm0.0006\pm0.00007$ &
  0.00279     &   --                     \\ \hline
$\Omega$        & 0.000856     &$0.00080\pm0.00025\pm0.0001$&
  0.000659    &$0.007\pm0.0035\pm0.00015$ \\ \hline
   \end{tabular}
\caption[]{Measured hadrons average multiplicities per hadronic event
compared to the fitted values. The first error is the experimental
one, the second error is due to uncertainty on hadron masses, widths
and branching ratios.}
\end{center}
\end{table}

\newpage

\begin{table}[ht]
   \begin{center}
   \begin{tabular}{| c || c | c || c | c |}       \hline
  & \multicolumn{2}{|c|}{$\sqrt{s} = 91.2$ GeV}  &
  \multicolumn{2}{|c|}{$\sqrt{s} = 29 \div 35$ GeV} \\ \hline
{\bf Hadrons}&  Calculated  & Measured        &   Calculated
 &   Measured      \\ \hline
$D^0$        &  0.231       & $0.221\pm0.012$ &   0.256
     & $0.225\pm0.035$ \\ \hline
$D^+$        &  0.0914      & $0.087\pm0.008$ &   0.106
     & $0.085\pm0.015$ \\ \hline
$D^{*+}$     &  0.107       & $0.088\pm0.0056$&   0.113
     & $0.192\pm0.036$ \\ \hline
$B$          &  0.0902      & $0.097\pm0.026$ &   0.0387
     & --              \\ \hline
$B^{*0}/B^0$ &  0.694       & $0.745\pm0.069$ &   0.697
     & --              \\ \hline
$\Lambda^+_c$&  0.0311      & $0.037\pm0.009$ &   0.0339
     & $0.055\pm0.025$ \\ \hline
$D_s$        &  0.0536      & $0.041\pm0.0076$&   0.0542
     & --              \\ \hline
   \end{tabular}
\caption[]{Measured heavy flavoured hadrons average multiplicities
per hadronic event compared to predictions of the model. The
experimental values\cite{hf} have been
averaged according to the procedure of ref.\cite{schm}.
Values of $R_c=0.17$, $R_b=0.22$ at $\sqrt{s} = 91.2$ and of
$R_c=0.36$, $R_b=0.09$ at $\sqrt{s} = 29 \div 35$ have been used.}
\end{center}
\end{table}

\begin{table}[hb]
   \begin{center}
   \begin{tabular}{| c || c | c |}
      \hline
  {\bf Parameter} &$\sqrt{s} =91.2$ GeV &$\sqrt{s}=29\div 35$ GeV
   \\ \hline\hline
Temperature(MeV)  &    2.06           &    1.08       \\ \hline
Volume(Fm$^3$)    &    1.34           &    0.47       \\ \hline
$\gamma_s$        &    0.017          &    0.018       \\ \hline
   \end{tabular}
\caption{Systematic uncertainties on fitted parameters.}
\end{center}
\end{table}


\begin{center}
\begin{figure}[htbp]
\mbox{\epsfig{file=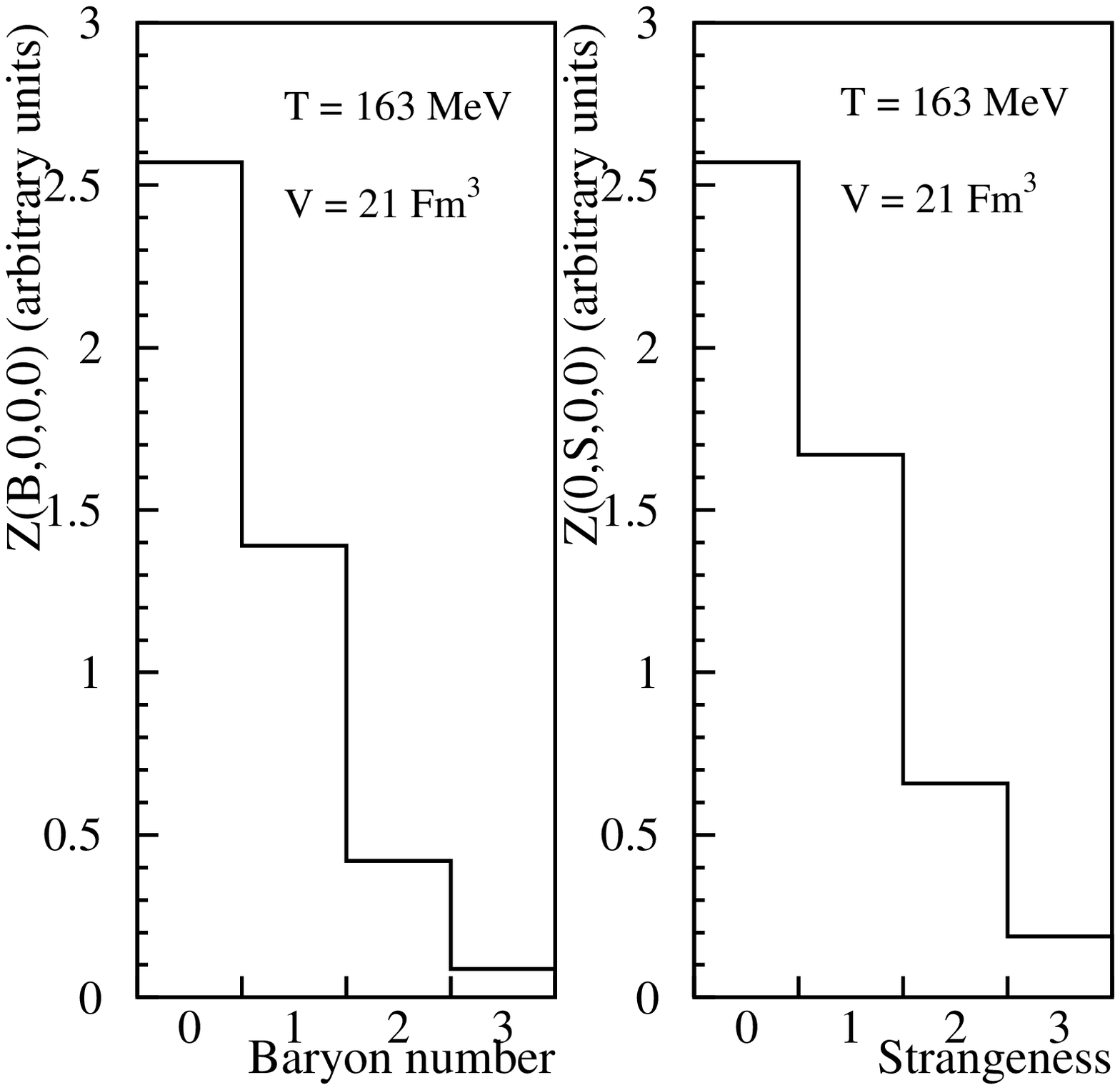,width=17cm}}
\caption{}
\end{figure}
\end{center}

\newpage

\begin{center}
\begin{figure}[htbp]
\mbox{\epsfig{file=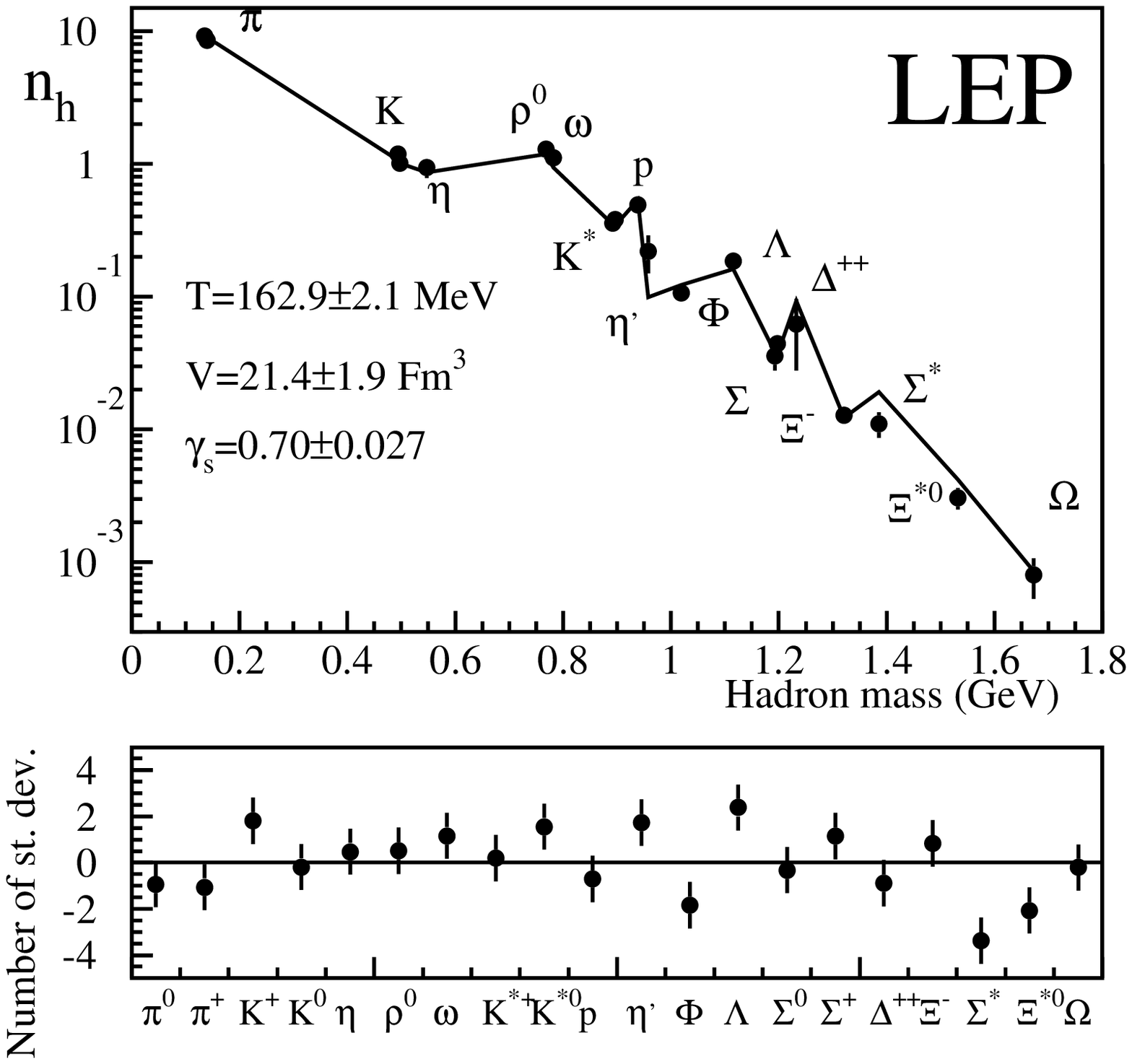,width=17cm}}
\caption{}
\end{figure}
\end{center}

\newpage

\begin{center}
\begin{figure}[htbp]
\mbox{\epsfig{file=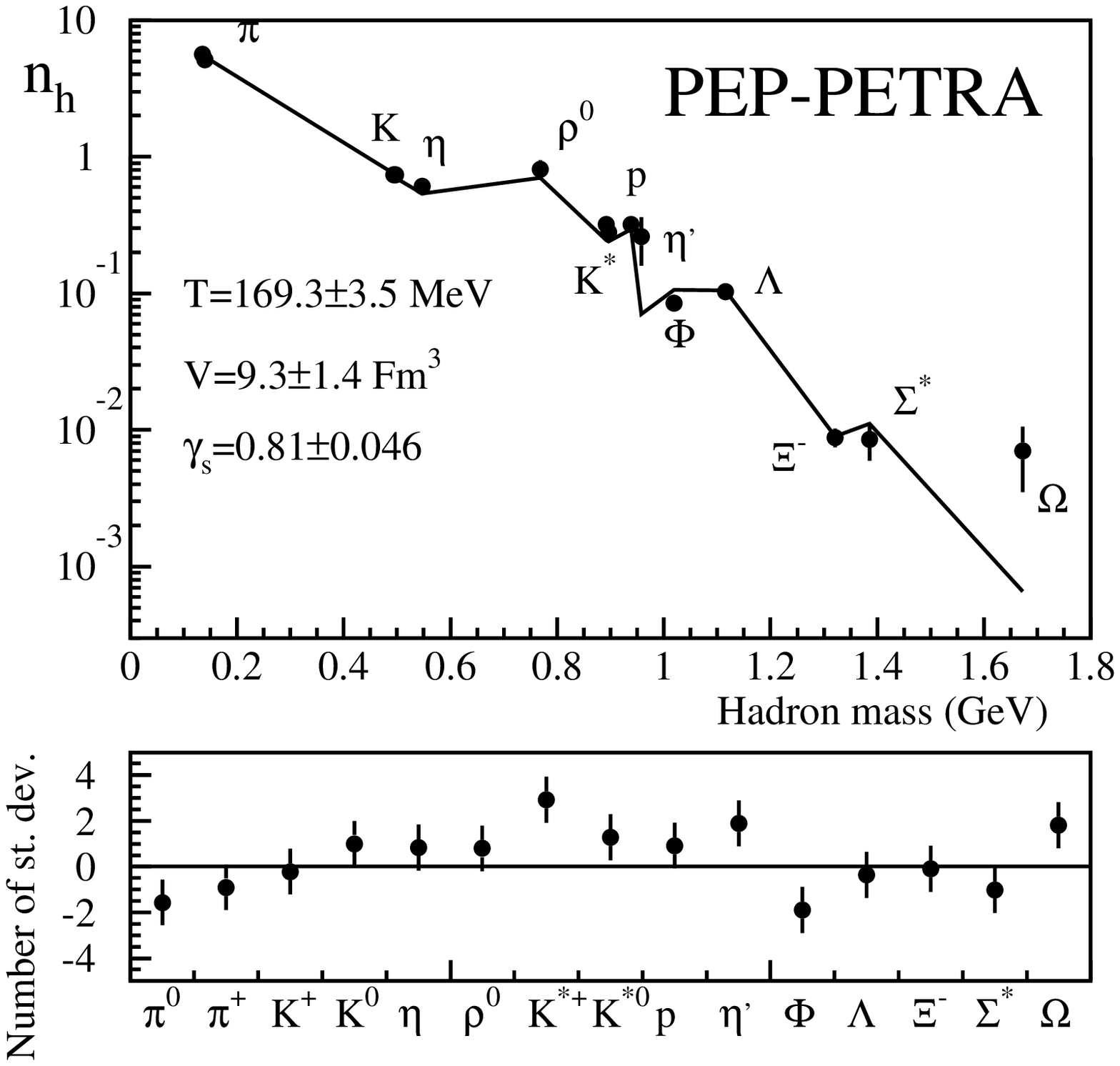,width=17cm}}
\caption{}
\end{figure}
\end{center}

\newpage

\begin{center}
\begin{figure}[htbp]
\mbox{\epsfig{file=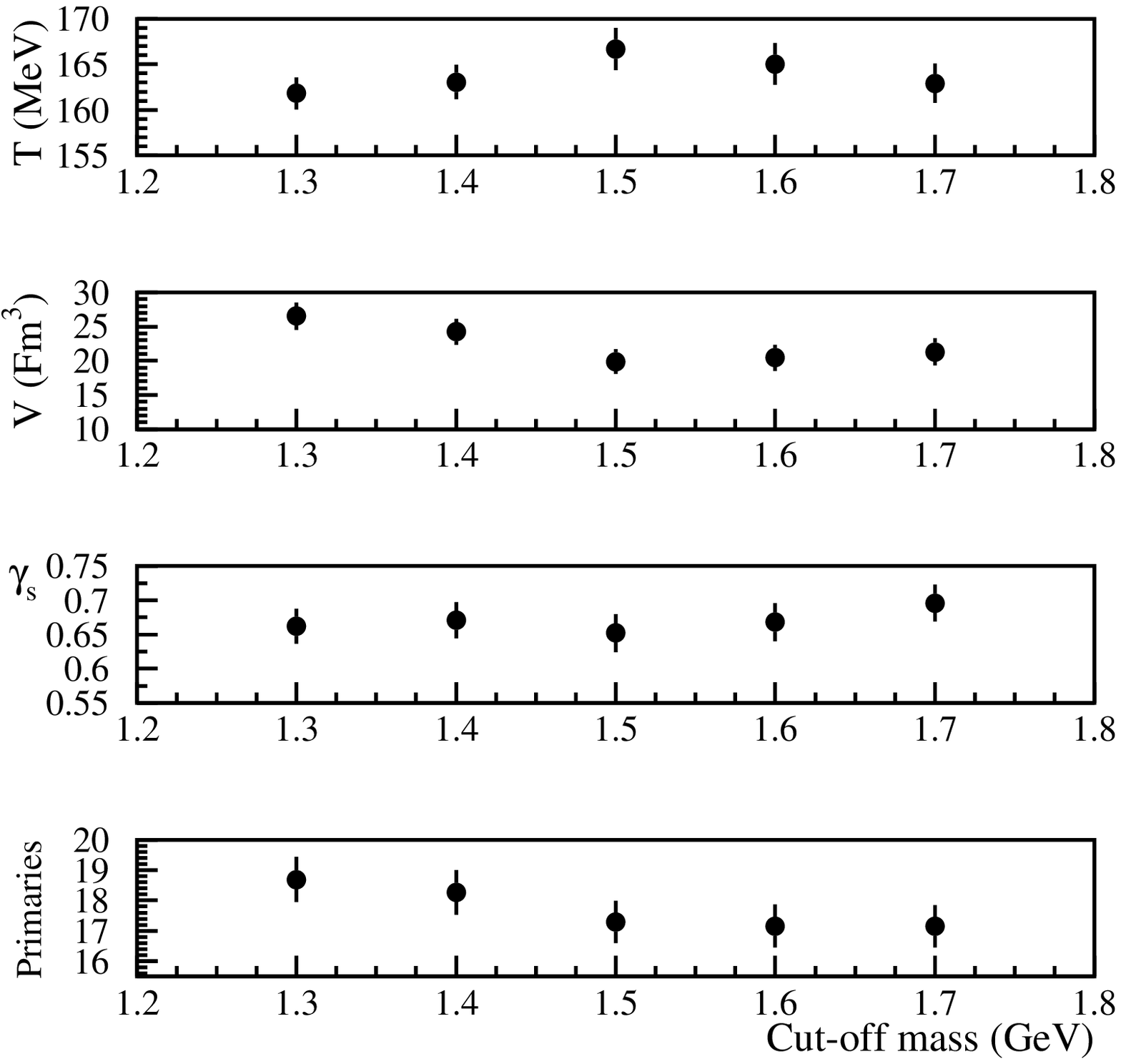,width=17cm}}
\caption{}
\end{figure}
\end{center}

\end{document}